\documentclass
[superscriptaddress,secnumarabic,amssymb,amsmath,nobibnotes,aps,prd,showkeys,showpacs,nofootinbib,onecolumn,notitlepage]{revtex4}%
\usepackage{graphicx}
\usepackage{epsf}
\usepackage{bm}
\usepackage{amsmath}
\usepackage{amsfonts}
\usepackage{amssymb}%
\providecommand{\U}[1]{\protect\rule{.1in}{.1in}}

\newcommand{\be}{\begin{equation}}
\newcommand{\ee}{\end{equation}}

\newcommand{\mincir}{\raise
-3.truept\hbox{\rlap{\hbox{$\sim$}}\raise4.truept\hbox{$<$}\ }}
\newcommand{\magcir}{\raise
-3.truept\hbox{\rlap{\hbox{$\sim$}}\raise4.truept\hbox{$>$}\ }}

\ifx\pdfoutput\relax\let\pdfoutput=\undefined\fi
\newcount\msipdfoutput
\ifx\pdfoutput\undefined\else
\ifcase\pdfoutput\else
\ifx\paperwidth\undefined\else
\ifdim\paperheight=0pt\relax\else\pdfpageheight\paperheight\fi
\ifdim\paperwidth=0pt\relax\else\pdfpagewidth\paperwidth\fi
\fi\fi\fi
\begin{document}
\title{Integrability from Point Symmetries in a family of Cosmological Horndeski
Lagrangians }
\author{N. Dimakis}
\email{nsdimakis@gmail.com}
\affiliation{Instituto de Ciencias F\'{\i}sicas y Matem\'{a}ticas, Universidad Austral de
Chile, Valdivia, Chile}
\author{Alex Giacomini}
\email{alexgiacomini@uach.cl}
\affiliation{Instituto de Ciencias F\'{\i}sicas y Matem\'{a}ticas, Universidad Austral de
Chile, Valdivia, Chile}
\author{Andronikos Paliathanasis}
\email{anpaliat@phys.uoa.gr}
\affiliation{Instituto de Ciencias F\'{\i}sicas y Matem\'{a}ticas, Universidad Austral de
Chile, Valdivia, Chile}
\affiliation{Institute of Systems Science, Durban University of Technology, PO Box 1334,
Durban 4000, Republic of South Africa}

\begin{abstract}
For a family of Horndeski theories, formulated in terms of a generalized
Galileon model, we study the integrability of the field equations in a
Friedmann-Lema\^{\i}tre-Robertson-Walker spacetime. We are interested in point
transformations which leave invariant the field equations. Noether's theorem
is applied to determine the conservation laws for a family of models that
belong to the same general class. The cosmological scenarios with or without
an extra perfect fluid with constant equation of state parameter are the two
important cases of our study. \ The De Sitter universe and ideal gas solutions
are derived by using the invariant functions of the symmetry generators as a
demonstration of our result. Furthermore, we discuss the connection of the
different models under conformal transformations while we show that when the
Horndeski theory reduces to a canonical field the same holds for the conformal
equivalent theory. Finally we discuss how singular solutions provides
nonsingular universes in a different frame and vice versa.

\end{abstract}
\maketitle
\date{\today}

\section{Introduction}

The plethora of phenomena which have been discovered the last few years have
led to the consideration of alternative/modified gravitational theories
\cite{clifton}. These extended theories of gravity provide additional terms in
the field equations which - in conjunction to those of General Relativity -
can explain the various phases of the universe. In physical science the
existence of a fundamental axiom, such as Hamilton's principle, is of
paramount importance. Among the theories that are generated by an action
integral, those that are of second-order in respect to their equations of
motion possess a distinguished role; from Newtonian Mechanics to General Relativity.

A particular family of theories which has drawn attention are the so-called
Horndeski theories. Horndeski, in 1974 \cite{hor} derived the most general
action for a scalar field in a four-dimensional Riemannian space in which the
Euler-Lagrange equations are at most of second-order. All relevant
scalar-tensor theories with this property, such as the Brans-Dicke
\cite{Brans,kofinasminas,adolfo}, the Galileon \cite{nik,gal02} and the
generalized Galileon \cite{mark} belong to the general family of Horndeski
theories. Although the latter are of second-order, they provide non-canonical
nonlinear field equations which - even for the simplest line element of the
underlying space - might not be able to lead to solutions in terms of
closed-form expressions. Due to the high level of nonlinearity, numerical
methods are applied in order to approximate the evolution of the system.
However, whether a solution actually exists is not always known. For that
reason, in this work, we are motivated to study the integrability of the field
equations for a class of families of Hordenski theories (or equivalently those
of the generalized Galileon Lagrangian \cite{Kobayashi}).

There are various methods to study the integrability of a system of
differential equations.\ Two of the most famous are: a) the existence of
invariant transformations, i.e. symmetries and b) the singularity analysis;
for a recent discussion and comparison of these two methods see
\cite{discsing}. Both of these have been applied widely in gravitational
theories \cite{aref0,aref1,aref2,aref3,aref4,aref5,aref6}. In several cases
the gravitational field equations, after a specific ansatz for the line
element is adopted, can be derived by point-like Lagrangians \cite{refmin}.
The application of Noether's theorem over the corresponding mini-superspace
action has been utilized for the determination of conservation laws/analytical
solutions in various models both at the classical as well as in the quantum
level, for instance see
\cite{nor0,nor0a,nor1,nor2,nor3,nor4,nor5,nor6,nor7,nor8,nor9,nor10} and
references therein. Noether's Theorem is the main mathematical tool that we
use in this study.

From the various different classes of generators of the invariant
transformations we consider the most simple; the one corresponding to the
so-called point symmetries. Point symmetries are the generators of the
invariant transformations in the base manifold in which the dynamical system
is defined. The application of Noether's theorem for point symmetries leads to
conservation laws linear in the momentum. Well known conservation laws of this
type from classical mechanics are those of the momentum and the angular
momentum. The plan of the paper it follows.

In Section \ref{themodel}, we present our model which is a special
consideration of the Hordenski Lagrangian and can be seen as a first
generalization of the canonical scalar-tensor theories. Moreover we consider
the cosmological scenario of an isotropic and homogeneous universe in which an
extra fluid exists with constant equation of state parameter. For that model
the field equations are calculated while the minisuperspace Lagrangian is
discussed. Furthermore our model admits four unknown functions which define
the specific form of the cosmological model. In Section \ref{symmetries}, we
apply Noether's theorem in the minisuperspace Lagrangian in order to specify
the unknown form of the functions which define the model and derive the
corresponding Noetherian conservation laws. For the completeness of our
analysis we consider separately the cases with or without an ideal gas and the
cases with zero or nonzero spatial curvature for the underlying spacetime. In
order to demonstrate the usefulness of our results we present some closed-form
(special) solutions in Section \ref{invariant}; which are derived from the
invariant functions of the admitted symmetry vectors. Finally in Section
\ref{discus} we discuss our results and specifically we discuss the relation
between our models under conformal transformations and we show how a singular
universe is mapped to a nonsingular universe under the change of the frame.

\section{The model}

\label{themodel}

The gravitational action integral that we consider is of the form
\begin{equation}
S=\int\!\!\sqrt{-g}\left[  h(\phi)R-\frac{\omega(\phi)}{2}\phi^{,\mu}%
\phi_{,\mu}-V(\phi)-\frac{g(\phi)}{2}\phi^{,\mu}\phi_{,\mu}\Box\phi\right]
d^{4}x +S_{m}, \label{action}%
\end{equation}
that falls into the class of a generalized Galileon (or Horndeski) model
\cite{Kobayashi,genlyGL}, where $S_{m}$ indicates any possible additional
matter content. Furthermore, for the background geometry, we consider a four
dimensional space of Lorentzian signature, isotropic and homogeneous with
spatially flat three dimensional geometry; that is, a
Friedmann-Lema\^{\i}tre-Robertson-Walker (FLRW) spacetime line element
\begin{equation}
ds^{2}=-N(t)^{2}dt^{2}+a(t)^{2}\left(  dx^{2}+dy^{2}+dz^{2}\right)  .
\label{FLRW}%
\end{equation}
The action integral (\ref{action}) is the simplest generalization of those
scalar-tensor theories where the kinetic energy is not coupled with functions
entailing derivatives of $\phi$. As we can see, when $g\left(  \phi\right)
\rightarrow0$, an action integral of that type, including the Brans-Dicke
case, is recovered.

Our goal is to derive all admissible gravitation models in \eqref{action} that
possess an integral of motion of the form previously described. In that
respect, note that the existence of $\omega(\phi)$ is not trivial since its
absorption with a reparameterization. $\phi=f(\varphi)=\int\omega
(\varphi)^{-1/2}d\varphi$ leads to the transformation of the Laplacian of
$\phi$ as (primes denote differentiation with respect to the argument)
\begin{equation}
\Box\phi=f^{\prime}(\varphi)\Box\varphi+f^{\prime\prime}(\varphi)\varphi
^{,\mu}\varphi_{,\mu},
\end{equation}
introducing a new kinetic energy squared term in action \eqref{action}. This
is in contrast to what happens in the case of a scalar field whose action does
not involve a coupling with derivatives of $\phi$ over its kinetic term, i.e.
when in our case $g\left(  \phi\right)  =0$, where without loss of generality
we can always select $\omega\left(  \phi\right)  $ to be a constant.

The mini-superspace Lagrangian that we can derive with the help of the
gravitation plus scalar field part of \eqref{action} and \eqref{FLRW} by
integrating out the spatial degrees of freedom reads\footnote{We assume that
the field $\phi$ inherits the symmetries of the FLRW spacetime (\ref{FLRW}).}
\begin{equation}
L\left(  N,a,\dot{a},\phi,\dot{\phi}\right)  =\frac{a^{2}g(\phi)\dot{a}%
\dot{\phi}^{3}}{N^{3}}-\frac{a^{3}g^{\prime}(\phi)\dot{\phi}^{4}}{6N^{3}%
}+\frac{6a^{2}h^{\prime}(\phi)\dot{a}\dot{\phi}}{N}+\frac{6ah(\phi)\dot{a}%
^{2}}{N}-\frac{a^{3}\omega(\phi)\dot{\phi}^{2}}{2N}+a^{3}NV(\phi),
\label{miniLag}%
\end{equation}
where the dot stands for differentiation with respect to time $t$.

In what regards the matter part, we start our investigation by considering the
$S_{m}=2\int\mathcal{L}_{m}d^{4}x$ contribution in \eqref{action} to be that
of a perfect fluid obeying the barotropic equation of state $P=\gamma\rho$,
with $P(t)$ and $\rho(t)$ the pressure and energy density respectively. In
order to perform the variation of $\mathcal{L}_{m}=\sqrt{-g}\rho$ with respect
to the metric we need to a priori define a continuity equation \cite{Schutz}.
We choose the following well known relation for an ideal gas in a FLRW
space-time:
\begin{equation}
\frac{\dot{\rho}}{\rho}=-3(1+\gamma)\frac{\dot{a}}{a}\rightarrow\rho
(t)=\rho_{0}a^{-3(\gamma+1)}, \label{continuityeq}%
\end{equation}
where $\rho_{0}$ is a constant of integration. Thus, the relevant addition to
the mini-superspace Lagrangian is
\begin{equation}
L_{m}\left(  N,a\right)  =\sqrt{-g}\rho(t)=N\rho_{0}a^{-3\gamma}.
\end{equation}
As a result the total Lagrangian reads
\begin{equation}
L_{tot}=L+2L_{m}=\frac{a^{2}g(\phi)\dot{a}\dot{\phi}^{3}}{N^{3}}-\frac
{a^{3}g^{\prime}(\phi)\dot{\phi}^{4}}{6N^{3}}+\frac{6a^{2}h^{\prime}(\phi
)\dot{a}\dot{\phi}}{N}+\frac{6ah(\phi)\dot{a}^{2}}{N}-\frac{a^{3}\omega
(\phi)\dot{\phi}^{2}}{2N}+a^{3}NV(\phi)+2N\rho_{0}a^{-3\gamma} \label{Lagtot}%
\end{equation}

It can be easily verified that the three Euler-Lagrange equations
\begin{subequations}
\label{Euler}%
\begin{align}
&  E_{N}=\frac{\partial L_{tot}}{\partial N}=0,\label{Eulercon}\\
&  E_{q}=\frac{\partial L_{tot}}{\partial q}-\frac{d}{dt}\left(
\frac{\partial L_{tot}}{\partial\dot{q}}\right)  =0,\quad\quad q=(a,\phi),
\end{align}
are completely equivalent to the field equations of motion of \eqref{action}
for the metric
\end{subequations}
\begin{align}
h(\phi)G_{\mu\nu}=  &  \frac{\omega}{2}\phi_{,\mu}\phi_{,\nu}-\frac{1}%
{2}g_{\mu\nu}\left(  \frac{\omega}{2}\phi^{,\kappa}\phi_{,\kappa}%
+V(\phi)\right)  -\frac{1}{2}g_{\mu\nu}G_{,\kappa}^{(1)}\phi^{,\kappa}%
+G_{(\mu}^{(1)}\phi_{,\nu)}\nonumber\\
&  -\frac{1}{2}G_{,X}^{(1)}\phi_{,\mu}\phi_{,\nu}\Box\phi+T_{\mu\nu}^{(m)}%
\end{align}
and the scalar field
\begin{equation}
h^{\prime}(\phi)R+\left(  \omega(\phi)\phi^{,\kappa}\right)  _{;\kappa}%
-\frac{\omega^{\prime}(\phi)}{2}\phi_{,\kappa}\phi^{,\kappa}-V^{\prime}%
(\phi)+\left(  G_{,X}^{(1)}\Box\phi\,\phi^{,\kappa}\right)  _{;\kappa
}+G_{,\phi}^{(1)}\Box\phi+\Box G^{(1)}=0,
\end{equation}
whenever the line-element \eqref{FLRW} is substituted and the isometries of
the space-time are inherited by the matter $\phi=\phi(t)$. The various
quantities that appear in these equations are: $G^{(1)}(\phi,X)=g(\phi)X$,
$X=-\frac{1}{2}\phi^{,\kappa}\phi_{,\kappa}$ and
\begin{equation}
T_{\mu\nu}^{(m)}=(\rho+p)u_{\mu}u_{\nu}+pg_{\mu\nu}%
\end{equation}
for the energy-momentum tensor of the fluid with $u^{\mu}=(\frac{1}{N},0,0,0)$
the comoving velocity.

\section{Symmetries and conservation laws}

\label{symmetries}

By considering the mini-superspace action to be form invariant under point
transformations generated by\footnote{For details on the Noether's theorem we
refer the reader to \cite{bluman}.}
\begin{equation}
Y=\chi(t,a,\phi,N)\frac{\partial}{\partial t}+\xi_{1}(t,a,\phi,N)\frac
{\partial}{\partial a}+\xi_{2}(t,a,\phi,N)\frac{\partial}{\partial\phi}%
+\xi_{3}(t,a,\phi,N)\frac{\partial}{\partial N}, \label{initgen}%
\end{equation}
one is naturally led to the well known infinitesimal criterion
\begin{equation}
\mathrm{pr}^{(1)}Y(L_{tot})+L_{tot}\frac{d\chi}{dt}=\frac{dF}{dt},
\label{symcrit}%
\end{equation}
where $\mathrm{pr}^{(1)}Y$ is the first prolongation of $Y$, i.e. the
extension of the generator in the first jet space spanned by $(t,a,\phi
,N,\dot{a},\dot{\phi},\dot{N})$. Clearly, since $\dot{N}$ does not appear in
\eqref{miniLag}, we can disregard the relevant term and just write
\begin{equation}
\mathrm{pr}^{(1)}Y=Y+\Phi_{1}\frac{\partial}{\partial\dot{a}}+\Phi_{2}%
\frac{\partial}{\partial\dot{\phi}},
\end{equation}
where $\Phi_{i}=\frac{d\xi_{i}}{dt}-\dot{q}^{i}\frac{d\chi}{dt}$, $q=(a,\phi
)$, $i=1,2$.

Application of \eqref{symcrit} leads to an overdetermined system of partial
differential equations to be solved for $\chi(t,a,\phi,N)$, $\xi_{i}%
(t,a,\phi,N)$, $i=1,...3$, and $F(t,a,\phi,N)$. The former is formed by
gathering the coefficients of all terms involving the derivatives of $a$,
$\phi$ and $N$, to which the functions entering the generator \eqref{initgen}
have no dependence. By gradually integrating the equations, we obtain the
following results (the basic steps to the solution can be found in appendix
\ref{appA}):

\begin{itemize}
\item An infinite dimensional symmetry group generated by
\begin{equation}
Y_{\infty}=\chi(t)\frac{\partial}{\partial t}-N\dot{\chi}(t)\frac{\partial
}{\partial N}, \label{Yinf}%
\end{equation}
with $\chi(t)$ an arbitrary function of time. Its appearance reflects the fact
that the mini-superspace action for \eqref{miniLag} is form invariant under
arbitrary time transformations. Its existence leads, trough Noether's second
theorem, to a differential identity between the Euler-Lagrange equations of
motion \eqref{Euler}. The latter implies the well known fact that not all of
them are independent, i.e. $L_{tot}$ is a constrained (or singular)
Lagrangian. It is well known that when an infinite dimensional symmetry group
is present (i.e. Noether's second theorem is applicable) the system is
necessarily singular, however the inverse is not true \cite{Sund}. In our case
this particular symmetry is a remnant of a more general group, the four
dimensional diffeomorphism, $\chi(x)^{\mu}\frac{\partial}{\partial x^{\mu}}$,
under which the full gravitational action \eqref{action} is form invariant.

\item Additionally to the previous group - which always exists for time
reparameterization invariant Lagrangians \cite{tchris} - we obtain a symmetry
generator if $h(\phi)$, $\omega(\phi)$, $g(\phi)$ and $V(\phi)$ satisfy
certain criteria. In particular, we see that if $\gamma\neq\frac{1}{3}$, then
\begin{equation}
Y=\frac{a}{3\gamma-1}\frac{\partial}{\partial a}-\frac{\phi}{\lambda}%
\frac{\partial}{\partial\phi}+N\frac{3\gamma}{3\gamma-1}\frac{\partial
}{\partial N}, \label{generator0a}%
\end{equation}
satisfies \eqref{symcrit} whenever
\begin{equation}
h(\phi)=\phi^{\frac{3(\gamma-1)\lambda}{3\gamma-1}},\quad g(\phi)=g_{0}%
\phi^{3(\lambda-1)},\quad V(\phi)=V_{0}\phi^{\frac{6(\gamma-1)\lambda}%
{3\gamma-1}-3\lambda},\quad\omega(\phi)=\omega_{0}\phi^{\frac{3(\gamma
-1)\lambda}{3\gamma-1}-2} \label{hvgw0a}%
\end{equation}
in which $\lambda$ is a constant. Note that the corresponding gauge function
appearing in \eqref{symcrit} is trivial in the calculation, i.e.
$F(t,a,\phi,N)=$const. and this remains that way for every case that we
examine later on in the analysis. Note also, that \eqref{generator0a} and
\eqref{hvgw0a} describe a particular solution of \eqref{symcrit} in the
special case when $\gamma=-1$. When the fluid contribution plays the role of a
cosmological constant, it can be absorbed inside the potential $V(\phi)$ (as
can be seen by the form of $L_{tot}$) and the general solution of the
$\gamma=-1$ case is the same as the one that we get if we completely omit the fluid.

On the other hand, for the special case where the perfect fluid describes
radiation, i.e. $\gamma=1/3$ the situation changes and the symmetry generator
assumes the general form
\begin{equation}
Y=a\frac{\tilde{g}(\phi)\tilde{g}^{\prime\prime}(\phi)}{2\tilde{g}^{\prime
}(\phi)^{2}}\frac{\partial}{\partial a}+\frac{\tilde{g}(\phi)}{\tilde
{g}^{\prime}(\phi)}\frac{\partial}{\partial\phi}+N\frac{\tilde{g}(\phi
)\tilde{g}^{\prime\prime}(\phi)}{2\tilde{g}^{\prime}(\phi)^{2}}\frac{\partial
}{\partial N}. \label{generator0b}%
\end{equation}
while the functions entering the action need to be
\begin{equation}
h(\phi)=\frac{1}{\tilde{g}^{\prime}(\phi)},\quad g(\phi)=g_{0}\frac{\tilde
{g}^{\prime}(\phi)^{3}}{\tilde{g}(\phi)^{3}},\quad V(\phi)=\frac{V_{0}}%
{\tilde{g}^{\prime}(\phi)^{2}},\quad\omega(\phi)=\omega_{0}\frac{\tilde
{g}^{\prime}(\phi)}{\tilde{g}(\phi)^{2}}-\frac{3\tilde{g}^{\prime\prime}%
(\phi)^{2}}{\tilde{g}^{\prime}(\phi)^{3}}, \label{hvgw0b}%
\end{equation}
where $\tilde{g}(\phi)$ is an arbitrary non-constant function. As a result,
for $\gamma=1/3$, there exists an infinite family of models belonging to the
general class of actions of the form \eqref{action} that possesses an integral
of motion of the type we are investigating.
\end{itemize}

\subsection{Particular case: $\rho_{0}=0$}

Let us see how the situation alters if we remove the ideal gas from our
considerations. In other words let us use the Lagrangian given by
\eqref{miniLag}, in the criterion \eqref{symcrit} instead of the Lagrangian
$L_{tot}$. Then, in addition to the diffeomorphism group characterized by
\eqref{Yinf} - which always exists for time reparameterization invariant
Lagrangians \cite{tchris} - we obtain a symmetry generator
\begin{equation}
Y=a\frac{(\lambda+3)\tilde{g}^{\prime}(\phi)^{2}-3\tilde{g}(\phi)\tilde
{g}^{\prime\prime}(\phi)}{6\tilde{g}^{\prime}(\phi)^{2}}\frac{\partial
}{\partial a}-\frac{\tilde{g}(\phi)}{\tilde{g}^{\prime}(\phi)}\frac{\partial
}{\partial\phi}+N\frac{(\lambda+3)\tilde{g}^{\prime}(\phi)^{2}-\tilde{g}%
(\phi)\tilde{g}^{\prime\prime}(\phi)}{2\tilde{g}^{\prime}(\phi)^{2}}%
\frac{\partial}{\partial N}, \label{generator1}%
\end{equation}
that satisfies \eqref{symcrit} when
\begin{equation}
h(\phi)=\frac{1}{\tilde{g}^{\prime}(\phi)},\quad g(\phi)=g_{0}\frac{\tilde
{g}^{\prime}(\phi)^{3}}{\tilde{g}(\phi)^{\lambda+6}},\quad V(\phi)=V_{0}%
\frac{\tilde{g}(\phi)^{\lambda+3}}{\tilde{g}^{\prime}(\phi)^{2}},\quad
\omega(\phi)=\frac{\omega_{0}\tilde{g}^{\prime}(\phi)}{\tilde{g}(\phi)^{2}%
}-\frac{3\tilde{g}^{\prime\prime}(\phi)^{2}}{\tilde{g}^{\prime}(\phi)^{3}}
\label{hvgw1}%
\end{equation}
with $\lambda$ being a constant and $\tilde{g}(\phi)$ again an arbitrary
(non-constant) function of $\phi$. Once more, we have an infinite set of
physically different models that are characterized by a function $\tilde
{g}(\phi)$. We can see that, in comparison to the $\gamma=1/3$ case, result
\eqref{hvgw1} is identical to \eqref{hvgw0b} in the special case when
$\lambda=-3$.

It is useful to study how the inclusion of spatial curvature (either positive
or negative) may alter the conditions under which a symmetry generator
appears. If we consider the line element
\begin{equation}
ds^{2}=-N^{2}dt^{2}+\frac{a(t)^{2}}{(1+\frac{k}{4}r^{2})^{2}}\left(
dx^{2}+dy^{2}+dz^{2}\right)  , \label{lineelk}%
\end{equation}
where $r^{2}=x^{2}+y^{2}+z^{2}$, Lagrangian \eqref{miniLag} is modified by the
addition of an extra term in the potential and reads
\begin{equation}
L_{k}=\frac{a^{2}g(\phi)\dot{a}\dot{\phi}^{3}}{N^{3}}-\frac{a^{3}g^{\prime
}(\phi)\dot{\phi}^{4}}{6N^{3}}+\frac{6a^{2}h^{\prime}(\phi)\dot{a}\dot{\phi}%
}{N}+\frac{6ah(\phi)\dot{a}^{2}}{N}-\frac{a^{3}\omega(\phi)\dot{\phi}^{2}}%
{2N}+a^{3}NV(\phi)-6\,k\,a\,Nh(\phi). \label{miniLag2}%
\end{equation}
Thus, for the same class of point transformations we considered earlier and
with the application of the infinitesimal symmetry criterion \eqref{symcrit},
it is straightforward to derive the result given by \eqref{generator1} and
\eqref{hvgw1} under the condition that $\lambda=-3$ or equivalently result
\eqref{generator0b} and \eqref{hvgw0b} corresponding to the radiation fluid
case where $\gamma=1/3$. In other words when $k\neq0$ we have the same
situation as in the $\gamma=1/3$ case. The function $\tilde{g}(\phi)$ once
more remains arbitrary, under the restriction of course of not being constant.
As we can see, even though the non vanishing of $k$ imposes a restriction
($\lambda=-3$) in comparison to the $k=0$ case, there still exists an infinite
number of models admitting an integral of motion of the type we consider in
this work.

Moreover, for the combined case where we have both a non-zero spatial
curvature and a perfect fluid, in other words when the Lagrangian is being
given by $L_{k}+2L_{m}$, it comes as no surprise that the existence of an
integral of motion implies that $\gamma=1/3$. In fact the result is once more
exactly the same with what we see in \eqref{generator0b} and \eqref{hvgw0b}.

Having derived the previous results, we can use Noether's second theorem and
derive the integrals of motion corresponding to each case. It can be easily
verified that
\begin{equation}
\frac{dI}{dt}=-\xi_{3}E_{0}+A^{i}E_{i},\quad i=1,2 \label{consevedq}%
\end{equation}
where the conserved charge is
\begin{equation}
I=\xi_{1}\frac{\partial L}{\partial\dot{a}}+\xi_{2}\frac{\partial L}%
{\partial\dot{\phi}} \label{charge}%
\end{equation}
while $A^{i}E_{i}$ denotes a linear combination of the two spatial equations
of motion. Of course, in place of $L$ there can be either $L_{tot}$ of
\eqref{Lagtot}, $L$ of $\eqref{miniLag}$ or $L_{k}$ of \eqref{miniLag2},
depending on the generator that we use and the case that we examine. In every
situation, we have on mass shell the corresponding conserved quantity
\eqref{charge} for each symmetry vector field $Y$. The fact that in every case
but the generic fluid with $\gamma\neq1/3$ an arbitrary function $\tilde
{g}(\phi)$ is involved in the generator as well as in the action itself,
implies that we possess an infinite collection of models - corresponding to
different sets of functions $h(\phi)$, $g(\phi)$, $V(\phi)$ and $\omega(\phi)$
- which admit at least one integral of motion of this type; hence being
integrable. Note here, that the fact that we are led to an infinite number of
models through the arbitrariness of $\tilde{g}(\phi)$ is owed to the use of
the reparameterization invariant Lagrangian \eqref{miniLag}. Had we chosen to
adopt the gauge $N=1$ at the Lagrangian level, then only a particular case
would have emerged where $\tilde{g}(\phi)=\phi^{\kappa}$, with the
corresponding $\chi(t)$ for this symmetry being $\chi(t)\propto t$. Of course,
due to the system being autonomous, $\partial_{t}$ also exists; the latter
being the only remnant of \eqref{Yinf} after fixing the gauge. The difference
appearing here in the arbitrariness of $\tilde{g}(\phi)$ lies in the
consideration of Lagrangian \eqref{miniLag}, which naturally generates the
constraint equation $E_{0}=\frac{\partial L}{\partial N}=0$ in \eqref{symcrit}
and \eqref{consevedq}, allowing the derivation of a larger class of symmetries.

The existence of an integral of motion of the form \eqref{charge} is of great
interest in the search of solutions. The system at hand possesses two degrees
of freedom that are bound by a constraint equation. Thus, the existence of $I$
implies that, in principle, we need only to solve the constraint equation
together with $I=$const. in order to fully integrate the system. As a result,
our problem is immediately reduced to one that involves only first order
differential equations. The usefulness of $I=$const. also lies in the fact
that it is linear in $\dot{a}$, an advantage which the constraint equation
does not have since it is non-linear in both velocities $\dot{a}$ and
$\dot{\phi}$. Nevertheless, even for the simplest of models defined by
$\tilde{g}(\phi)$, the situation can be highly complicated. In most of cases
it may be more useful if the result is used in a way to distinguish any
existing solutions that are invariant under the action of the generator, i.e.
derive the characteristic of $Y$ for a given model $\tilde{g}(\phi)$ so as to
deduce a possible relation between $a$ and $\phi$ and use the latter together
with the integral of motion and the constraint equation.

\section{Invariant solutions}

\label{invariant}

In what follows, and in order to demonstrate the importance of our results, we
present a few illustrative applications of invariant solutions that we derive
in the gauge $N=1$ with and without a perfect fluid. For a treatment of how
the conserved quantity can be utilized to lead to more general solutions
expressed in an arbitrary gauge we refer the reader to appendix \ref{appB}.
Note that the solutions which we present here are derived with a mentality of
keeping all terms inside the action, i.e. we do not express any additional
solutions which may have a vanishing $V_{0}$, $g_{0}$ or $\omega_{0}$.

\begin{enumerate}
\item \label{casesol1} \textbf{Perfect fluid, $\gamma\neq1/3$ case}. Let as
consider the FLRW spacetime with zero spatial curvature, i.e. $k=0$. As is
known invariant solutions can be constructed with the help of the symmetry
generator. For example, in the case of a generic perfect fluid with
$\gamma\neq1/3$, generator \eqref{generator0a} implies an invariant relation
of the form $a^{\frac{1-3 \gamma}{\lambda}} \phi=$const. Hence, it is easy to
derive the solution
\begin{equation}
\label{sol01}a(t)=t^{\gamma/3},\quad\phi(t)=\phi_{0} t^{\frac{3 \gamma-1}{3
\gamma\lambda}}%
\end{equation}
with
\begin{equation*}%
\begin{split}
g_{0}  &  = -\frac{3 \gamma^{2} \lambda^{2} \phi_{0}^{\frac{6 \gamma\lambda
}{1-3 \gamma}} \left[ 6 \lambda^{2} \left( 3 \gamma\left( 2 \gamma\rho_{0}
\phi_{0}^{-\frac{3 (\gamma-1) \lambda}{3 \gamma-1}}+\gamma V_{0} \phi
_{0}^{\frac{6 \gamma\lambda}{1-3 \gamma}}-2\right) +4\right) +(1-3 \gamma)^{2}
\omega_{0}\right] }{(3 \gamma-1)^{3} (3 \gamma(\lambda-1)-3 \lambda+1)}\\
\omega_{0}  & = \frac{6 \lambda^{2} \left[ 3 \gamma\left( \gamma(1-3 \gamma)
\rho_{0} \phi_{0}^{-\frac{3 (\gamma-1) \lambda}{3 \gamma-1}}+2 \gamma V_{0}
\phi_{0}^{\frac{6 \gamma\lambda}{1-3 \gamma}}+2\right) -4\right] }{(1-3
\gamma)^{2}} .
\end{split}
\end{equation*}
The on mass shell value of the conserved quantity $I$ for solution
\eqref{sol01} is $I=\frac{6 \gamma(\gamma+1) \rho_{0}}{3 \gamma-1}$.

It is interesting to note what occurs when the theory assumes a
Brans-Dicke-like form, i.e. when $h(\phi)=\phi$ and $\omega(\phi) = \omega_{0}
\phi^{-1}$. This happens when $\lambda= \frac{3 \gamma-1}{3 (\gamma-1)}$ and
it leads to a particular solution with $\gamma=-1$ (cosmological constant
contribution in the action) that reads
\begin{equation}
\label{sol01BD}a(t)=t^{-1/3},\quad\phi(t)=\phi_{0} t^{2}%
\end{equation}
where now $\omega_{0}=-5/3$, $V_{0}= -2 \rho_{0}$, while $g_{0}$ is kept
arbitrary. We can observe that solution \eqref{sol01BD} satisfies the same
condition as the vacuum Brans-Dicke solution (for a spatially flat universe),
namely that $a(t)^{3} \phi(t) \propto t$ (in the gauge $N=1$) \cite{Chauvet}.
This is explained by the fact that for $\gamma=-1$ the potential in
\eqref{hvgw0a} becomes a constant with the value $V_{0}= -2 \rho_{0}$ so that
it effectively cancels the cosmological constant contribution of the fluid in \eqref{Lagtot}.

Nevertheless, for an arbitrary $\gamma$, we can see how the symmetry that is
present in the Brans-Dicke model - satisfying $a^{6} \phi$=constant - is
generalized in the class of models we are considering here. In particular we
encounter the more general relation $a^{\frac{1-3 \gamma}{\lambda}} \phi
=$constant, that is implied by the existing symmetry generator. Of course in
our case, \eqref{sol01} is not a general solution of the equations of motion
but a particular one, which however, as a power-law, is of special
cosmological interest.

\item \label{casesol2} \textbf{In the absence of a fluid, $\tilde{g}%
(\phi)=e^{-\mu\phi}$ case}. For a spatially flat FLRW line-element, we choose
the function appearing in \eqref{hvgw1} to be $\tilde{g}(\phi)=e^{-\mu\phi}$.
After an appropriate redefinition of the constants $\omega_{0}$, $g_{0}$ and
$V_{0}$, we derive the corresponding model to be
\begin{equation}
h(\phi)=e^{\mu\phi},\quad\omega(\phi)=\omega_{0}e^{\mu\phi},\quad
g(\phi)=g_{0}e^{\mu(\lambda+3)\phi},\quad V(\phi)=V_{0}e^{-\mu(\lambda+1)\phi
}.\label{func1}%
\end{equation}
A special solution in terms of a power-law for the scale factor exists
\begin{equation}
a(t)=t^{\sigma},\quad\phi(t)=\frac{2}{\mu(\lambda+2)}\ln(\phi_{0}%
t),\quad\lambda\neq-2,\label{sol1}%
\end{equation}
when the constants $\omega_{0}$ and $V_{0}$ are given by
\begin{equation*}%
\begin{split}
\omega_{0} &  =\frac{2g_{0}\phi_{0}^{2}((\lambda+2)(3\sigma-1)-2)}{\mu
(\lambda+2)^{2}}+\mu^{2}((\lambda+2)(\lambda+3)\sigma+\lambda),\\
V_{0} &  =\frac{4g_{0}\phi_{0}^{4}(3\sigma-1)+2\mu^{3}\phi_{0}^{2}%
(\lambda+2)((\lambda+2)\sigma+1)(3(\lambda+2)\sigma-\lambda)}{\mu^{3}%
(\lambda+2)^{3}},
\end{split}
\end{equation*}
while the rest parameters remain free. It can be seen that on this solution
the conserved quantity $I$ (the Noetherian conservation law) becomes zero.

\item \label{casesol3} \textbf{Case $\tilde{g}(\phi)=\phi^{-\mu}$, without a
perfect fluid}. Again in \eqref{hvgw1}, we make the choice $\tilde{g}%
(\phi)=\phi^{-\mu}$ and obtain (once more with an appropriate redefinition of
the constants $V_0, g_0, \omega_0$ in the action and $\lambda\rightarrow \lambda/\mu$) the subsequent set of functions entering the action
\begin{equation}
h(\phi)=\phi^{\mu+1},\quad\omega(\phi)=\omega_{0}\phi^{\mu-1},\quad
g(\phi)=g_{0}\phi^{3(\mu-1)+\lambda},\quad V(\phi)=V_{0}\phi^{2-\mu
-\lambda}.\label{func2}%
\end{equation}
Two solutions for which the conserved quantity is again zero can be easily
derived for this model.

The first solution is again a power-law
\begin{equation}
a(t)=t^{\sigma},\quad\phi(t)=\phi_{0}t^{\kappa}, \label{sol2}%
\end{equation}
where
\begin{align*}
\lambda &  =\frac{2}{\kappa }-2 \mu +1,\\
\omega_{0}  &  =-g_0 \kappa ^2 (\mu -1) \phi_0^{2/\kappa }+ g_0 \kappa  (3 \sigma -1) \phi_0^{2/\kappa }+\frac{4 \sigma }{\kappa ^2}+\frac{2 (\mu +1) (\sigma +1)}{\kappa }-2 (\mu +1)^2,\\
V_{0}  &  =\frac{1}{2} \phi_0^{2/\kappa } \left(g_0 \kappa ^4 \phi_0^{2/\kappa }+ g_0 \kappa ^3 (3 \sigma -1) \phi_0^{2/\kappa }+2 \kappa ^2 (\mu +1)^2+2 \kappa  (\mu +1) (5 \sigma -1)+4 \sigma  (3 \sigma -1)\right),
\end{align*}
are the constants appearing in \eqref{func2}.

Once more we can also distinguish a Brans-Dicke sub-case that appears when
$\mu=0$. It is interesting to observe that condition $a(t)^{3} \phi(t) \propto t$ of
the vacuum Brans-Dicke model does not apply here, although it was still
relevant in the corresponding model of case \ref{casesol1}. As we can see, the
power $\sigma$ in the scale factor expression is not connected to the power of
time in the $\phi(t)$. Only if we choose to turn the action into the pure
Brans-Dicke form does \eqref{sol2} with $\mu=0$ satisfy the aforementioned condition: By
setting $g_{0}=0$ and $V_{0}=0$, which immediately results in $\kappa=1-3\sigma$ or $\kappa = -2 \sigma$.

Another solution we can derive for the $\tilde{g}(\phi)=\phi^{-\mu}$ model
leads to a space-time of a constant scalar Ricci curvature, a de Sitter
universe that is:
\begin{equation}
a(t)=e^{\sigma t},\quad\phi(t)=\phi_{0}e^{\kappa t} \label{sol3}%
\end{equation}
when the following relations hold
\begin{align*}
\lambda &  =1-2 \mu,\\
\omega_{0}  &  = g_0 \kappa  \left(\kappa (1-\mu) +3 \sigma \right)-\frac{2 (\mu +1) \left(\kappa  (\mu +1) -\sigma \right)}{\kappa },\\
V_{0}  &  =\frac{g_0 \kappa ^4}{2}+\frac{3}{2} g_0 \kappa ^3 \sigma +\kappa ^2 (\mu +1)^2+5 \kappa  (\mu +1) \sigma +6 \sigma^2.
\end{align*}

\item \label{casesol4} \textbf{Case $k\neq 1$, $\tilde{g}(\phi)=e^{-\mu\phi}$, without a
perfect fluid}. Let us furthermore consider the case with nonzero spatial curvature $k\neq
0$. Then, from the two particular types of solutions we examined above in cases \ref{casesol2} and \ref{casesol3}, only for
$\tilde{g}(\phi)=e^{-\mu\phi}$ arises the following configuration:
\begin{equation}%
\begin{split}
&  a(t)=t,\quad\phi(t)=-\frac{2}{\mu}\ln\left(  \phi_{0}\,t\right) \\
\omega_{0}=k  &  -8g_{0}\mu\phi_{0}^{2},\quad V_{0}=4\mu\phi_{0}^{2}\left(
2g_{0}\mu\phi_{0}^{2}-k\right)  ,
\end{split}
\end{equation}
which, as long as the scale factor is concerned, it corresponds to the $k=-1$
solution of Einstein's gravity in vacuum, that is the Milne solution. Here, it
is given for any sign of $k$ with the difference being carried in the coupling
(through $\omega_{0}$ and $V_{0}$) that is necessary for its existence. The
integral of motion - calculated with the help of \eqref{charge},
\eqref{miniLag2} and \eqref{generator0b} - assumes on mass shell the value
$I=\frac{2\,k}{\mu\phi_{0}^{2}}$.

\end{enumerate}

As is obvious, we are able - just by choosing a particular function $\tilde
{g}(\phi)$ - to find through \eqref{hvgw1} or \eqref{hvgw0b} the corresponding
class of integrable models that admit a conserved quantity of the
aforementioned form. As we already stated, it is due to the existence of the
constraint equation \eqref{Eulercon} that one can in principle obtain the
general solution by considering only the first order system $E_{0}%
=\frac{\partial L}{\partial N}=0$ and $I=$const. However, because of the
complexity of the equations this is a highly nontrivial task.

\section{Discussion}

\label{discus}

We remark that for the cases without the extra matter fluid where we have
obtained an infinite set of models owed to the arbitrariness of $\tilde
{g}(\phi)$, there exists an interesting connection among them. These models
can be mapped to each other by conformal transformations. Take for example two
different models whose action is characterized by two different scalar field
functions $\tilde{g}_{1}(\phi_{1})$ and $\tilde{g}_{2}(\phi_{2})$. Then, the
transformation
\begin{equation}
\phi_{1}=\tilde{g}_{1}^{-1}\circ\tilde{g}_{2}(\phi_{2}),\quad g_{\mu\nu}%
^{(1)}=\frac{1}{(\tilde{g}_{1}^{-1}\circ\tilde{g}_{2})^{\prime}(\phi_{2}%
)}g_{\mu\nu}^{(2)}, \label{conf1}%
\end{equation}
where $g_{\mu\nu}^{(1)}$ and $g_{\mu\nu}^{(2)}$ are the two corresponding
spacetimes, maps the one action to the other. What is more, if we try to map
the relevant form of the action to the Einstein frame we can see that an
interesting \textquotedblleft degeneration" occurs\footnote{A similar
situation can be seen to arise in the result of \cite{nor1}, where again
infinite models in a family of scalar tensor theories was recovered. By means
of conformal transformations they can be mapped in the Einstein frame to a
case of a minimally coupled scalar field with exponential potential.}: The
different models that we get for the various $\tilde{g}(\phi)$ all collapse to
a single action. Take for example the case where the initial action is
characterized by the functions of $\phi$ given by \eqref{hvgw1}. It can be
seen to be mapped into an action of the form
\begin{equation}
\bar{S}=\int\!\!\sqrt{-\bar{g}}\left[  \frac{1}{2}\bar{R}-\frac{1}{2}%
\Phi^{,\mu}\Phi_{,\mu}-V_{1}e^{\sigma\Phi}-e^{-\sigma\Phi}\left(  \omega
_{1}\Phi^{,\mu}\Phi_{,\mu}\bar{\Box}\Phi+\omega_{2}(\Phi^{,\mu}\Phi_{,\mu
})^{2}\right)  \right]  d^{4}x \label{actein}%
\end{equation}
by a conformal transformation in the metric and a re-parametrization of
$\phi(\Phi)$. The new to the old variables are related by the expressions
\begin{equation}
\bar{g}_{\mu\nu}=\frac{2}{\tilde{g}^{\prime}(\phi)}g_{\mu\nu},\quad\Phi
(\phi)=\frac{1}{\sigma}\ln\left(  \frac{V_{0}\tilde{g}(\phi)^{\lambda+3}%
}{4V_{1}}\right)  \label{conf2}%
\end{equation}
with $V_{1}$ a constant and $\sigma$, $\omega_{1}$ and $\omega_{2}$ in
\eqref{actein} being related to the initial constants $g_{0}$, $V_{0}$,
$\omega_{0}$ and $\lambda$. We can see that the difference of this action lies
in the existence of a term that is quadratic to the kinetic energy of the
scalar field, while in \eqref{action} we considered a theory that has at most
linear expressions in $X=-\frac{1}{2}\phi^{,\mu}\phi_{\mu}$.

However, what is also interesting here is that when $V\left(  \Phi\right)
\simeq e^{\sigma\Phi}$ dominates then the contribution of the terms
$e^{-\sigma\Phi}\left(  \omega_{1}\Phi^{,\mu}\Phi_{,\mu}\bar{\Box}\Phi
+\omega_{2}(\Phi^{,\mu}\Phi_{,\mu})^{2}\right)  \simeq0$, which means that
(\ref{actein}) reduces to the Action Integral of a minimally coupled scalar
field. The latter is exactly the limit which relates a canonical scalar field
between the Jordan and the Einstein frames.

In general, great care needs to be taken when using conformal transformations.
The fact that all the different models for the various choices of $\tilde
{g}(\phi)$ can be mapped to \eqref{actein} does not make them equivalent to
the latter, or even to each other (due to \eqref{conf1}) for that matter. Two
actions in order to be physically equivalent, they need to be mapped by gauge
transformations of the theories under consideration. In the case of
gravitational actions, those are the four dimensional diffeomorphism of
space-time. As we know, conformal transformations cannot always be attributed
to coordinate changes. As a result, the gravitational space-time arising in
each situation is generally different.

In order to see that consider the nonsingular solution \eqref{sol3}
corresponding to a de Sitter universe, which came out of a model characterized
by $\tilde{g}(\phi)=\phi^{-\mu}$. If we choose to map this model to
\eqref{actein} then, by virtue of \eqref{conf2} we obtain
\begin{equation}
\bar{N}(t)=-\frac{2\phi(t)^{\mu+1}}{\mu}=c_{1}\,e^{\kappa(\mu+1)t},\quad
\bar{a}(t)=-\frac{2\phi(t)^{\mu+1}}{\mu}e^{\sigma t}=c_{1}\,e^{[\kappa
(\mu+1)+\sigma]t},
\end{equation}
where $\bar{N}$, $\bar{a}$ are the new lapse and scale factor respectively,
while $c_{1}$ is a constant. It is easy to see that if you go from this
metric
\begin{equation}
ds^{2}=-\bar{N}^{2}dt^{2}+\bar{a}^{2}(dx^{2}+dy^{2}+dz^{2})
\end{equation}
to the gauge where $\bar{N}=1$ (so as to compare with what we have in
\eqref{sol3}) by performing the time transformation
\begin{equation}
\bar{N}(t)dt=d\tau\Rightarrow t=\frac{\ln\left(  \frac{\kappa(\mu+1)\tau
}{c_{1}}\right)  }{\kappa(\mu+1)}%
\end{equation}
we derive, with the appropriate scalings in $x$, $y$ and $z$, the line
element
\begin{equation}
ds^{2}=-d\tau^{2}+\tau^{\psi}(dx^{2}+dy^{2}+dz^{2}),
\end{equation}
where $\psi$ is a constant. What was an exponential solution with a constant
Ricci scalar in the theory we are investigating, has now become a power law
with $R(\tau)\propto\tau^{-2}$ and a space-time with curvature singularity at
$\tau=0$ in the system described by action \eqref{actein}. Hence we can see
that, whenever the conformal transformation does not correspond to a general
coordinate transformation, the gravitational properties of the system are
bound to change and the solutions represent different geometries. For
discussions on the relation between analytical solutions and physical
quantities between the Jordan and the Einstein frames see
\cite{frame1,frame2,grgts} and references therein.

Additionally to the previous geometrical considerations, extra care needs to
be taken when a fluid is used as a matter source. This is owed to the fact
that you need to pre-define a continuity equation in order to derive a rule
for the variation of the energy density $\rho$ with respect to the metric
$g_{\mu\nu}$. In the beginning of our analysis we considered a perfect fluid
that is characterized by continuity equation \eqref{continuityeq}. It is a
well known fact that the latter is not conformally invariant, while its
solution is necessarily utilized at the level of the minisuperspace
Lagrangian. Henceforth, after a conformal transformation is being made, one
needs to take into account a different fluid, which is now interacting with
the scalar field, in order to make such a correspondence possible. This also
results in a change of the physical behaviour of the system, since the
properties of the matter source need to be altered.

In a future work we plan to study the integrability of other Horndeski
theories by including more terms in the action integral and for other kind of
transformations which leaves the field equations invariants, such as the
generalized symmetries as also to investigate the effects of the conformal
transformations in Hordenski theories.

\appendix

\section{Calculation of the symmetry generator}

\label{appA}

As indicated in the main text, application of \eqref{symcrit} leads to an
overdetermined system of partial differential equations. The latter is formed
by gathering and demanding that are zero the coefficients multiplying terms
involving derivatives of $a$, $\phi$ and $N$. For Lagrangian $L_{tot}$ as
written in \eqref{Lagtot} we infer from the coefficients of $\dot{a}^{3}$,
$\dot{a}^{2} \dot{N}$ and $\dot{a}^{2} \dot{\phi}$ that we need to set
respectively
\begin{equation}
\label{condchi}%
\begin{cases}
\partial_{a} \chi & = 0\\
\partial_{N} \chi & = 0\\
\partial_{\phi}\chi & = 0
\end{cases}
\Rightarrow\chi=\chi(t).
\end{equation}
The fourth order coefficients $\dot{a}^{2} \dot{\phi}^{2}$ and $\dot{a}
\dot{\phi}^{2} \dot{N}$ each imply
\begin{equation}
\partial_{a} \xi_{2} = 0 \quad\text{and} \quad\partial_{N} \xi_{2} = 0,
\end{equation}
which means that $\xi_{2}=\xi_{2}(t,\phi)$. However, with the help of
$\chi=\chi(t)$, we can get a further restriction from the coefficient of
$\dot{a}\dot{\phi}^{2}$ that leads to
\begin{equation}
\label{xi2phi}\partial_{t} \xi_{2} = 0 \Rightarrow\xi_{2} = \xi_{2}(\phi).
\end{equation}
Due to \eqref{condchi} and \eqref{xi2phi} , the terms involving $\dot{\phi
}^{3} \dot{N}$ and $\dot{\phi}^{3}$ bring about the conditions
\begin{equation}
\label{xi1aphi}%
\begin{cases}
\partial_{N} \xi_{1} & = 0\\
\partial_{t} \xi_{1} & = 0
\end{cases}
\Rightarrow\xi_{1}= \xi_{1} (a,\phi).
\end{equation}
Thanks to the above restrictions \eqref{condchi}, \eqref{xi2phi} and
\eqref{xi1aphi} we also get
\begin{equation}
\label{condF1}%
\begin{cases}
\partial_{N} F & = 0\\
\partial_{\phi}F & = 0\\
\partial_{a} F & = 0
\end{cases}
\Rightarrow F=F(t)
\end{equation}
from $\dot{N}$, $\dot{\phi}$ and $\dot{a}$ respectively.

The equation emanating from the coefficient of $\dot{a}^{2}$ can be solved
algebraically with respect to $\xi_{3}(t,a,\phi,N)$
\begin{equation}
\xi_{3}(t,a,\phi,N) = N \left( 2 \partial_{a} \xi_{1}(a,\phi) + \frac{\xi
_{1}(a,\phi)}{a}+\frac{\xi_{2}(\phi) h^{\prime}(\phi)}{h(\phi)}-\dot{\chi
}(t)\right) ;
\end{equation}
the latter being a consequence of the fact that no derivative of $N$ enters in
the Lagrangian. After this step one can see that the equation extracted from
the coefficient of $\dot{a} \dot{\phi}^{3}$ is
\begin{equation}
a h(\phi) \xi_{2}(\phi) g^{\prime}(\phi)-g(\phi) \left[ h(\phi) \left( 5 a
\partial_{a} \xi_{1} (a,\phi)+\xi_{1}(a,\phi)-3 a \xi_{2}^{\prime}%
(\phi)\right) +3 a \text{$\xi$211}(\phi) h^{\prime}(\phi)\right]  =0
\end{equation}
and it can be immediately integrated to yield
\begin{equation}
\xi_{1} (a,\phi) = \frac{1}{6} a \left( \frac{\xi_{2}(\phi) g^{\prime}(\phi
)}{g(\phi)}-\frac{3 \xi_{2}(\phi) h^{\prime}(\phi)}{h(\phi)}+3 \xi_{2}%
^{\prime}(\phi)\right) +\frac{\xi_{0}(\phi)}{a^{1/5}} .
\end{equation}

With the dependence with respect to $a$ being now completely specified, the
relation produced by the coefficient of $\dot{a} \dot{\phi}$ involves only
unknown functions of $\phi$. Hence, we can now start gathering coefficients
with respect to powers of $a$ that appear inside it. With the help of a useful
reparametrization $g(\phi)= g_{1}^{\prime}(\phi)^{3}$ one can
straightforwardly obtain:
\begin{align}
\xi_{2}(\phi)  &  = \frac{c_{1}+c_{2} g_{1}(\phi)}{g_{1}^{\prime}(\phi)}\\
\xi_{0}(\phi)  &  = \frac{c_{3}}{h(\phi)^{3/2}},
\end{align}
where the $c_{i}$'s indicate constants of integration. The only appearance of
time in the remaining coefficient is inside the zero-th order component (that
multiplies no derivative of $\dot{a}$, $\dot{\phi}$ or $\dot{N}$) and it
implies that
\begin{equation}
F(t) = \text{const}.
\end{equation}
Thus, the gauge function becomes trivial.

With the results up to here all dependence of the functions of the generator
with respect to $a$, $t$ and $N$ is specified. Inside each of the remaining
equations we can gather coefficients with respect to $a$, since now the
unknown functions involve only $\phi$. However, due to the existence of
$\gamma$ the way that the coefficients are to be gathered depends on its
value. We can distinguish two main cases:

\begin{itemize}

\item Case $\gamma\neq1/3$. The equation produced by the zero-th order
coefficient leads to
\begin{align}
V(\phi)  &  = \frac{c_{4} h(\phi)^{2}}{\left( c_{1} + c_{2} g_{1}(\phi)\right)
^{3}}\\
h(\phi)  &  = c_{5} \left(  c_{1}+ c_{2} g_{1}(\phi)\right) ^{\frac
{3(\gamma-1)}{3\gamma-1}}\\
c_{3}  &  =0 .\nonumber
\end{align}
We note that, since we want the most general result, we avoid any special
solution that leads to vanishing of any of the functions involved in our
starting action. The $\dot{\phi}^{2}$ component leads to
\begin{equation}
\omega(\phi) = c_{6} \left(  c_{1}+c_{2} g_{1}(\phi)\right) ^{\frac{1+3\gamma
}{1-3\gamma}}g_{1}^{\prime2}%
\end{equation}
while the $\dot{\phi}^{4}$ gives rise to the third order equation
\begin{equation}
\label{thirdord1}g_{1}^{\prime\prime\prime} g_{1}^{\prime} g_{1} -2
(g_{1}^{\prime\prime})^{2} g_{1} + g_{1}^{\prime\prime} (g_{1}^{\prime})^{2}=0
\end{equation}
which under a transformation $g_{1} = \exp\left( \int g_{2} (\phi)d\phi\right)
$ becomes
\begin{equation}
\label{thirdord2}g_{2}^{\prime\prime} g_{2} -2 (g_{2}^{\prime})^{2}=0
\end{equation}
with general solution $g_{2} = \frac{c_{8}}{\phi+c_{9}}$. Thus, the final
function needed so that all equations are satisfied is
\begin{equation}
\label{thirdord3}g_{1}(\phi)= c_{7} (\phi+ c_{9})^{c_{8}}.
\end{equation}
At this point the system of equations is completely satisfied. By absorbing
the trivial constant $c_{9}$ inside $\phi$ with a translation and with an
appropriate parametrization of the rest of the constants we obtain result
\eqref{generator0a}, \eqref{hvgw0a}. Additionally, we can observe that the
function $\chi(t)$ remained arbitrary through the calculation and this
explains the existence of \eqref{Yinf}.

\item Case $\gamma= 1/3$. With this choice of $\gamma$ the zero-th order
equation implies
\begin{align}
V(\phi)  &  = \frac{c_{4} h(\phi)^{2}}{\left( c_{1} + c_{2} g_{1}(\phi)\right)
^{3}}\\
c_{2} =  &  c_{3} = 0 .\nonumber
\end{align}
The relations from the $\dot{\phi}^{2}$ and $\dot{\phi}^{4}$ coefficients
respectively lead to
\begin{align}
\omega(\phi)  &  = c_{5} h(\phi) (g_{1}^{\prime})^{3}- \frac{h^{\prime}%
(\phi)^{2}}{h(\phi)}\\
h (\phi)  &  = c_{7} \frac{e^{c_{6} g_{1}(\phi)}}{g_{1}^{\prime}(\phi)} .
\end{align}
It is a matter of reparametrizing the constants of integration and an
introduction of a new function $\tilde{g}(\phi)$ as $g_{1}(\phi)= \frac
{1}{\tilde{g}(\phi)}$ to obtain result \eqref{generator0b}, \eqref{hvgw0b} and
of course the same comments as in the previous case, for the arbitrariness of
$\chi(t)$, hold.
\end{itemize}

For the sake of completeness we have to point out that apart from the case
$\gamma=1/3$ there are also other values of $\gamma$ that can lead to a
different gathering of terms in the coefficients of powers of $a$; namely
$\gamma=-1$ and $\gamma=-7/5$. The first gives the same result as the case
without fluid (see eqs. \eqref{generator1}, \eqref{hvgw1}) with the sole
difference of a potential that is $V(\phi)= V_{0} \frac{\tilde{g}^{\lambda+3}%
}{\tilde{g}^{\prime}}- 2\rho_{0}$ in place of the one appearing in
\eqref{hvgw1}. That is one that cancels the cosmological constant role of the
fluid. The second case, $\gamma=-7/5$, after appropriate reparametrizations
leads to the exact same result as the generic $\gamma\neq1/3$ result and that
is why we do not make a separate presentation of it here.

\section{More general solutions}

\label{appB}

In section \ref{invariant} we presented some invariant solutions where the
functional dependence between $a$ and $\phi$ can be extracted with the help of
the infinitesimal symmetry generator. Although these are particular solutions
which can be derived in a rather simple manner, they are the most
cosmologically interesting. Even though we have proven the existence of an
integral of motion $I$, thus reducing the problem to solving two first order
differential equations: the constraint and $I=$const. The acquirement of the
general solution is still a very difficult task, especially due to the
non-linearity of the constraint equation in both derivatives involved $\dot
{a}$ and $\dot{\phi}$.

Here, we want to exhibit a method with which the integral of motion $I$ can be
used to derive the solution for the spatially flat case in the absence of a
perfect fluid, when $I=0$. Some of the solution we derived in section
\ref{invariant} lead to $I=0$ but still they are not the general solution of
this case, but rather particular solutions of it. Although this method does
not lead to the full solution (where $I=$const.) of the system, it can be
applied for an arbitrary function $\tilde{g}(\phi)$. Thus, giving in closed
form the full solution with the property $I=0$ for any model of the integrable
type we are considering.

At first let us construct the conserved quantity $I$ of \eqref{charge} by
using the $\xi_{1}$ and $\xi_{2}$ of \eqref{generator1}
\begin{equation}
\xi_{1} = a\frac{(\lambda+3)\tilde{g}^{\prime}(\phi)^{2}-3\tilde{g}%
(\phi)\tilde{g}^{\prime\prime}(\phi)}{6\tilde{g}^{\prime}(\phi)^{2}}, \quad
\xi_{2} = -\frac{\tilde{g}(\phi)}{\tilde{g}^{\prime}(\phi)}%
\end{equation}
and make the transformation $a(t)= \exp\left( \int u(t) dt \right) $. Then
relation $I=0$ reduces to
\begin{equation}%
\begin{split}
u \left( 4 (\lambda+3) N^{2} \tilde{g}(\phi)^{\lambda+6}\tilde{g}^{\prime
}(\phi)-6 g_{0} \tilde{g}(\phi) \dot{\phi}^{2} \tilde{g}^{\prime}(\phi
)^{4}\right)  + \dot{\phi} \Big[  &  \tilde{g}^{\prime}(\phi)^{2} \left( 2
\omega_{0} N^{2} \tilde{g}(\phi)^{\lambda+5} - g_{0} (\lambda+7) \dot{\phi
}^{2} \tilde{g}^{\prime}(\phi)^{3} \right) \\
& + \tilde{g}(\phi) \tilde{g}^{\prime\prime}(\phi) \left( 3 g_{0} \dot{\phi
}^{2} \tilde{g}^{\prime}(\phi)^{3} - 2 (\lambda+3) N^{2} \tilde{g}%
(\phi)^{\lambda+5} \right) \Big] = 0,
\end{split}
\end{equation}
which can be algebraically solved with respect to the function $u(t)$.
Substitution of the latter inside the constraint equation $\frac{\partial
L}{\partial_{N}} = 0$ leads to
\begin{equation}
\label{algN}%
\begin{split}
8 (\lambda+3)^{2} V_{0} \tilde{g}(\phi)^{4 (\lambda+5)} N^{8} -4 \dot{\phi
}^{2} \left( 3 \omega_{0}^{2}-\omega_{0} (\lambda+3)^{2}+6 g_{0} (\lambda+3)
V_{0}\right)  \tilde{g}(\phi)^{3 (\lambda+5)} \tilde{g}^{\prime}(\phi)^{3}
N^{6}\\
+ 2 g_{0} \dot{\phi}^{4} \left( 6 \omega_{0} (\lambda+7)+9 g_{0} V_{0}-2
(\lambda+6) (\lambda+3)^{2}\right)  \tilde{g}(\phi)^{2 (\lambda+5)} \tilde
{g}^{\prime}(\phi)^{6} N^{4}\\
+ 3 g_{0}^{2} (-3 \omega_{0}+\lambda(\lambda+2)-19) \dot{\phi}^{6} \tilde
{g}(\phi)^{\lambda+5} \tilde{g}^{\prime}(\phi(t))^{9} N^{2} + 9 g_{0}^{3}
\dot{\phi}^{8} \tilde{g}^{\prime}(\phi)^{12} =0,
\end{split}
\end{equation}
which is an eighth order algebraic equation with respect to $N(t)$. However,
only even powers of $N$ appear. So, by making a substitution $N(t)=\sqrt
{n(t)}$ in \eqref{algN}, the latter reduces to a fourth order polynomial
equation and can be solved analytically with respect to $n(t)$. We refrain for
giving the expression here, but general formulas can be found in the
bibliography \cite{Abramowitz}. As a result we are able, in the case of $I=0$,
to derive the solution of the system in a purely algebraic manner for every
admissible function $\tilde{g}(\phi)$ without even fixing the gauge, since we
have not set $N=1$ or any of the rest two degrees of freedom to be a specific
function of $t$. We only satisfied dynamical equations, even though it was just for the special case $I=0$. The solution is expressed through the relations
\begin{equation}
N(t)= \sqrt{n(\phi,\dot{\phi})}, \quad a(t)= e^{\int\!\! u(\phi,\dot{\phi})
dt}%
\end{equation}
in terms of an arbitrary function $\phi(t)$, i.e. the scalar field plays effectively
the role of time.

\begin{acknowledgments}
This work is financial supported by FONDECYT grants 3150016 (ND), 1150246 (AG)
and 3160121 (AP). AP thanks the Durban University of Technology for the
hospitality provided while part of this work was performed.
\end{acknowledgments}


\begin{thebibliography}{99}                                                                                               %


\bibitem {clifton}T. Clifton, P.G. Ferreira, A. Padilla and C. Skordis, Phys.
Rep. \textbf{513,} 1 (2012),

\bibitem {hor}G.W. Horndeski, Int. J. Ther. Phys. \textbf{10,} 363 (1974)

\bibitem {Brans}C. Brans and R.H. Dicke, Phys. Rev. \textbf{124}, 195 (1961)

\bibitem {kofinasminas}G. Kofinas and M. Tsoukalas, EPJC \textbf{76,} 686 (2016)

\bibitem {adolfo}A. Cisterna, T. Delsate and M. Rinaldi, Phys. Rev. D
\textbf{92,} 044050 (2015)

\bibitem {nik}A. Nicolis, R. Rattazzi and E. Trincherini, Phys.\ Rev. D
\textbf{79,} 064036 (2009)

\bibitem {gal02}C.~Deffayet, G.~Esposito-Farese and A.~Vikman, Phys.\ Rev.\ D
\textbf{79}, 084003 (2009)

\bibitem {mark}M. Trodden and K. Hinterbichler, Class. Quant. Grav.
\textbf{28,} 204003 (2011)

\bibitem {Kobayashi}T. Kobayashi, M. Yamaguchi and J. Yokoyama,Prog. Theor.
Phys. \textbf{126,} 511 (2011)

\bibitem {discsing}A. Paliathanasis and P.G.L. Leach, Int. J. Geom. Meth. Mod.
Phys. \textbf{13,} 1630009 (2016)

\bibitem {aref0}S. Cotsakis, P.G.L. Leach and H. Pantazi, Grav. Cosm. 4, 314 (1998)

\bibitem {aref1}S.D. Majaraj and P.G.L Leach, J. Math. Phys. \textbf{37}, 430 (1996)

\bibitem {aref2}A.M. Msomi, K.S. Govinder and S.D.\ Maharaj, Int. J. Theor.
Phys. \textbf{51}, 1290 (2012)

\bibitem {aref3}S. Cotsakis, J. Demaret, Y. De Rop and L. Querella,
Phys.\ Rev. D \textbf{48}, 4595 (1993)

\bibitem {aref4}S. Cotsakis and P.G.L. Leach, J. Phys. A: Math. Gen.
\textbf{27}, 1625 (1994)

\bibitem {aref5}A. Paliathanasis, J.D. Barrow and P.G.L. Leach, Phys. Rev. D
\textbf{94}, 023525 (2016)

\bibitem {aref6}A. Paliathanasis and P.G.L. Leach, Phys. Lett. A \textbf{380},
2815 (2016)

\bibitem {refmin}R.S. Palais, Commun. Math. Phys. \textbf{69,} 19 (1979)

\bibitem {nor0}S. Capozziello, R. De Ritis, C. Rubano and P. Scudellaro, Riv.
Nuovo Cim. \textbf{19,} 1 (1996)

\bibitem {nor0a}S. Capozziello, E. Piedipalumbo, C. Rubano and P. Scudellaro,
Phys. Rev. D. \textbf{80,} 104030 (2009)

\bibitem {nor1}Petros A. Terzis, N. Dimakis and T. Christodoulakis, Phys. Rev.
D \textbf{90}, 123543 (2014)

\bibitem {nor2}Y. Zhang, Y.-G. Gong, Z.-H. Zhu, Phys. Lett. B \textbf{688}, 13 (2010)

\bibitem {nor3}B. Vakili, Phys. Lett. B \textbf{664}, 16 (2008)

\bibitem {nor4}B. Vakili, Phys. Lett. B \textbf{738}, 488 (2014)

\bibitem {nor5}H. M. Sadjadi, Phys. Lett. B \textbf{718}, 270 (2012)

\bibitem {nor6}A. Paliathanasis, Class. Quantum Gravit. \textbf{33}, 075012 (2016)

\bibitem {nor7}A. Zampeli, T. Pailas, Petros A. Terzis and T. Christodoulakis,
JCAP \textbf{16}, no.05 066 (2016)

\bibitem {nor8}M.F. Shamir and M. Ahmad, EPJC \textbf{77,} 55 (2017)

\bibitem {nor9}K. Atazadeh and F.\ Darabi, EPJC \textbf{72,} 2016 (2012)

\bibitem {nor10}H. Dong, J. Wang and X. Meng, EPJC \textbf{73,} 2543 (2013)

\bibitem {genlyGL}G. Leon and E.N. Saridakis, JCAP \textbf{13}, no. 03 025 (2013)

\bibitem {Schutz}B. F. Schutz, Phys. Rev. D \textbf{2}, no. 12 2762 (1970)

\bibitem {bluman}G.W. Bluman and S. Kumei, Symmetries of Differential
Equations, (1989) (Springer-Verlag, New York)

\bibitem {Sund}K. Sundermeyer, Constrained Dynamics, Lect. Notes Phys. Vol.
169, Springer-Verlag (1982)

\bibitem {tchris}T. Christodoulakis, N. Dimakis and Petros A. Terzis, J. Phys.
A: Math. Theor. \textbf{47}, 095202 (2014)

\bibitem {Chauvet}P. Chauvet, Astrophysics and Space Science \textbf{90},
51-58 (1983)

\bibitem {frame1}V. Faraoni and E. Gunzig, Int. J. Theor. Phys. \textbf{38},
217 (1999)

\bibitem {frame2}M. Postma and M. Volponi, Phys.\ Rev. D \textbf{90}, 103516 (2014)

\bibitem {grgts}M. Tsamparlis, A. Paliathanasis, S. Basilakos and S.
Capozziello, \ Gen. Rel. Grav. \textbf{45}, 2003 (2013)

\bibitem {Abramowitz}M. Abramowitz and I. A. Stegun, Handbook of Mathematical
Functions with Formulas, Graphs, and Mathematical Tables, New York: Dover (1972)
\end{thebibliography}
\end{document}